\documentclass[twocolumn,aps,prb,showpacs]{revtex4}

\usepackage{epsfig}
\usepackage{graphicx}
\usepackage{amsmath}
\usepackage{bm}

\begin{document}

\title{Impurities, Quantum Interference and Quantum Phase Transitions in $s$-wave superconductors}
\author{Dirk K.~Morr and Jaesung Yoon}
\affiliation{Department of Physics, University of Illinois at
Chicago, Chicago, IL 60607}
\date{\today}

\begin{abstract}
We study the effects of quantum interference in impurity structures
consisting of two or three magnetic impurities that are located on
the surface of an $s$-wave superconductor. By using a
self-consistent Bogoliubov-de Gennes formalism, we show that quantum
interference leads to characteristic signatures not only in the
local density of states (LDOS), but also in the spatial form of the
superconducting order parameter. We demonstrate that the signatures
of quantum interference in the LDOS are qualitatively, and to a
large extent quantitatively unaffected by the suppression of the
superconducting order parameter near impurities, which illustrates
the robustness of quantum interference phenomena. Moreover, we show
that by changing the interimpurity distance, or the impurities'
scattering strength, the $s$-wave superconductor can be tuned
through a series of first order quantum phase transitions in which
the spin polarization of its ground state changes. In contrast to
the single impurity case, this transition is not necessarily
accompanied by a $\pi$-phase shift of the order parameter, and can
in certain cases even lead to its enhancement. Our results
demonstrate that the superconductor's LDOS, its spin state, and the
spatial form of the superconducting order parameter are determined
by a subtle interplay between the relative positions of the
impurities and their scattering strength.

\end{abstract}

\maketitle

\section{Introduction}

Over the last few years, quantum interference effects arising from
scattering of conduction electrons by multiple impurities have
attracted significant experimental
\cite{Man00,Der02,Chi02,Hol01,Bra02,Pie04,Gross04} and theoretical
\cite{Corral,Fla00,Morr03a,Morr03b,Morr04,UCSC,Sta05} attention. For
example, Manoharan {\it et al.}~\cite{Man00} used a corral of
magnetic impurities on the surface of a metallic host to demonstrate
that quantum interference and the resulting formation of eigenmodes
can be used to create quantum images (also called {\it quantum
mirages}). Complementary to this experiment, Derro {\it et
al.}~\cite{Der02}, using scanning tunneling spectroscopy (STS),
reported the formation of novel resonance states arising from
impurities in the one-dimensional chains of YBa$_2$Cu$_3$O$_{6+x}$.
Quantum interference effects have also been studied in optical
quantum corrals \cite{Chi02}, in quantum dots \cite{Hol01}, in
triangular corrals \cite{Bra02}, in ferromagnetic islands
\cite{Pie04}, and around molecules \cite{Gross04}.

A number of theoretical studies have focused on quantum interference
effects in metallic host systems \cite{Corral}, as well as
conventional \cite{Fla00,Morr03a,Morr03b,Morr04} and unconventional
\cite{UCSC,Sta05} superconductors. Superconducting (SC) host systems
with $s$-wave symmetry are of particular interest for the
investigation of quantum interference effects for two reasons.
First, a magnetic impurity induces a fermionic bound state in an
$s$-wave superconductor, which can be used as the object (i.e., the
"quantum candle") for the formation of a quantum image
\cite{Morr04}, while a non-magnetic impurity does not. This
qualitative difference between a magnetic and non-magnetic impurity
allows one to study quantum interference effects separately from the
formation of fermionic impurity states, in contrast to
unconventional superconductors \cite{UCSC,Sta05,Sal97,Balatsky_RMP}.
Second, quantum interference can be used to tune an $s$-wave
superconductor through a series of first order quantum phase
transition in which the spin polarization of the superconductor's
ground state changes \cite{Morr03b,Sal97,Balatsky_RMP,Sak70,Baz00}.
Quantum interference phenomena are therefore not only of great
fundamental interest, but might also possess important applications
in the field of spin electronics \cite{Wolf01} and quantum
information technology \cite{Kane98}, as these phenomena might lead
to the creation of new types of quantum qubits.

In this article, we study quantum interference phenomena arising in
structures consisting of two or three magnetic impurities that are
located on the surface of an $s$-wave superconductor. We employ a
self-consistent Bogoliubov-de Gennes (BdG) formalism \cite{Gen89}
which allows us to study the effects of quantum interference not
only on the formation of impurity bound states, but also on the
spatial form of the superconducting order parameter. We obtain three
important results. First, we show that quantum interference leads
not only to characteristic signatures in the local density of states
(LDOS) but also in the spatial form of the superconducting order
parameter. In particular, when the distance between impurities is
changed, the order parameter at the sites of the impurities (i.e.,
the {\it on-site} order parameter) exhibits Friedel-like
oscillations. These oscillations mirror the ones in the frequency of
the hybridized impurity bound state. Second, we find that the
characteristic signatures of quantum interference in the LDOS, such
as, e.g., the formation of bonding and antibonding bound states
around two magnetic impurities, and the oscillation of the bound
state frequencies with interimpurity distance \cite{Fla00,Morr03b},
remain qualitatively and to a large extent quantitatively unchanged
by the suppression of the order parameter around impurities. This
results demonstrate the robustness of quantum interference
phenomena. The physical origin of this robustness lies in the fact
that, while the superconducting order parameter is suppressed by
impurities, it recovers its bulk value over a length scale which is
set by the Fermi wavelength, $\lambda_F$, and not by the
superconducting coherence length, $\xi_c$, where for many $s$-wave
superconductors typically $\xi_c \gg \lambda_F$. Third, we show that
quantum interference leads to a series of first order quantum phase
transitions, in which the spin polarization of the superconductor's
ground state changes. These transitions can be achieved by
increasing the impurities' scattering strength, or, which is
experimentally more relevant, by changing the interimpurity
distance. While this type of quantum phase transition is well
understood for a single impurity \cite{Sak70}, we find that the
phase transitions associated with multiple impurities exhibit
qualitatively new features. In particular, we show that in contrast
to the single impurity case, these transitions are not necessarily
accompanied by a $\pi$-phase shift of the on-site order parameter,
and can, in certain cases, even lead to its enhancement. The
theoretical predictions for the spatial form of the superconducting
order parameter discussed below are of particular importance in
light of recent experimental progress in developing Josephson STS
\cite{JSTS}, which suggests that our predictions can be tested in
the near future.

The rest of the paper is organized as follows. In
Sec.\ref{BdGtheory} we briefly introduce the BdG formalism. In
Secs.~\ref{2impurities} and \ref{3impurities} we present our results
for the case of a two and three impurity systems, respectively.
Finally, in Sec.~\ref{conclusion} we summarize our results and
conclusions.

\section{Bogoliubov-de Gennes Formalism}
\label{BdGtheory}

In order to treat quantum interference between electrons that are
scattered by multiple impurities and a spatial variation of the
superconducting order parameter on equal footing, we use the
self-consistent Bogoliubov-de Gennes (BdG) formalism \cite{Gen89}.
Within this formalism, one solves the eigenvalue equation
\begin{equation}
\sum_{\bf j}
\begin{pmatrix}
 H^+_{\bf ij} & \Delta_{\bf ij} \\
\Delta^*_{\bf ij} & -H^-_{\bf ij}
\end{pmatrix}
\begin{pmatrix}
u_{{\bf j},n} \\
v_{{\bf j},n}
\end{pmatrix}
= E_n
\begin{pmatrix}
u_{{\bf i},n} \\
v_{{\bf i},n}
\end{pmatrix}
\label{EVeq}
\end{equation}
with $H^\pm_{\bf ij}=t_{\bf ij}+(\pm J_{\bf i} S_{\bf i}-\mu)
\delta_{\bf ij}$. Here, $t_{\bf ij}$ is the hopping integral between
sites ${\bf i}$ and ${\bf j}$, and $\mu$ is the chemical potential.
Below, we consider a two-dimensional system, in which only the
nearest-neighbor hopping integral $t$ and the next-nearest neighbor
hopping integral $t^\prime$ are non-zero. For concreteness, we take
$t^\prime/t=0.2$ and $\mu/t=-2$, leading to an almost circular Fermi
surface (FS) with $k_F \approx \pi/2$. $J_{\bf i}$ and $S_{\bf i}$
are the magnetic scattering potential and the spin of the impurity
located at site ${\bf i}$, respectively. Below, we consider for
concreteness magnetic impurities with spin $S=1/2$. In an $s$-wave
superconductor, the superconducting order parameter $\Delta_{\bf
ij}=\Delta_{\bf i} \delta_{\bf ij}$ is local and given by
\begin{equation}
\Delta_{\bf i}=-V \sum_n u_{{\bf i},n} v_{{\bf i},n} \tanh\left(
\frac{E_n}{2 k_B T} \right) \ , \label{SCgap}
\end{equation}
where $V$ is the effective pairing interaction and the sum runs over
all eigenstates of the system. Eqs.(\ref{EVeq}) and (\ref{SCgap})
are solved self-consistently in order to obtain the energy, $E_n$,
of all eigenstates of the system, the local superconducting order
parameter, $\Delta_{\bf i}$, as well as the Bogoliubov coefficients
$u_{{\bf i},n}, v_{{\bf i},n}$ of state $n$ at site ${\bf i}$.
Below, we take $V/t=2.5$, such that in the clean system, the
translationally invariant superconducting order parameter is given
by $\Delta_{\bf i}/t \approx 0.112$ . Finally, the local density of
states at site ${\bf i}$ is obtained via
\begin{equation}
N(\omega, {\bf i})=\sum_n \left[ u^2_{{\bf i},n} \delta(\omega-E_n)
+ v^2_{n,{\bf i}} \delta(\omega+E_n) \right] \ .
\end{equation}
Unless otherwise stated, we consider in the following a system with
$N=N_x \times N_y$ sites where $N_x=28$ and $N_y=18$. This choice of
$N_x,N_y$ accounts for the spatial anisotropy of the impurity
structures considered below.

Note that within the BdG-formalism, one assumes that the impurity
spin is a classical, static variable, which corresponds to the limit
$JS=const.$~and $S \rightarrow \infty$. In a fully gaped s-wave
superconductor, this approximation is justified even when
considering magnetic impurities with $S<\infty$, and in particular
small values of $S$. Specifically, it was shown \cite{Sat92} that no
Kondo effect occurs in a fully gapped $s$-wave superconductor for
sufficiently small coupling between the magnetic impurities and the
delocalized electrons. Hence, it is not necessary to consider the
full quantum dynamics of the magnetic impurity, which can thus be
treated as a classical, static variable. This finding is also
supported by the experimentally measured LDOS around a single
magnetic impurity in an $s$-wave superconductor \cite{Yaz97}. These
experiments have reported the existence of two peaks inside the
superconducting gap, which are induced by the magnetic impurity.
Within the self-consistent BdG-formalism or the non-self-consistent
${\hat T}$-matrix approach \cite{Sal97,Balatsky_RMP,Shiba68} these
peaks are a direct consequence of the static nature of the magnetic
impurity and are the spectroscopic signature of a fermionic (Shiba)
bound state.

Finally, within the BdG-formalism, any interaction between the
magnetic impurities is only taken into account to the extent that it
determines the angle, $\alpha$, between the direction of the
impurity spins. Within a ${\hat T}$-matrix approach, it was shown in
Ref.~\cite{Morr03b} that for two impurities, the cases $\alpha=0$
and $\alpha \not =0$ are qualitatively very similar \cite{com2}. We
therefore focus below on the case of impurities with parallel spin,
i.e., $\alpha=0$.

\section{Two magnetic impurities}
\label{2impurities}

%
%
\begin{figure}[t]
\epsfig{file=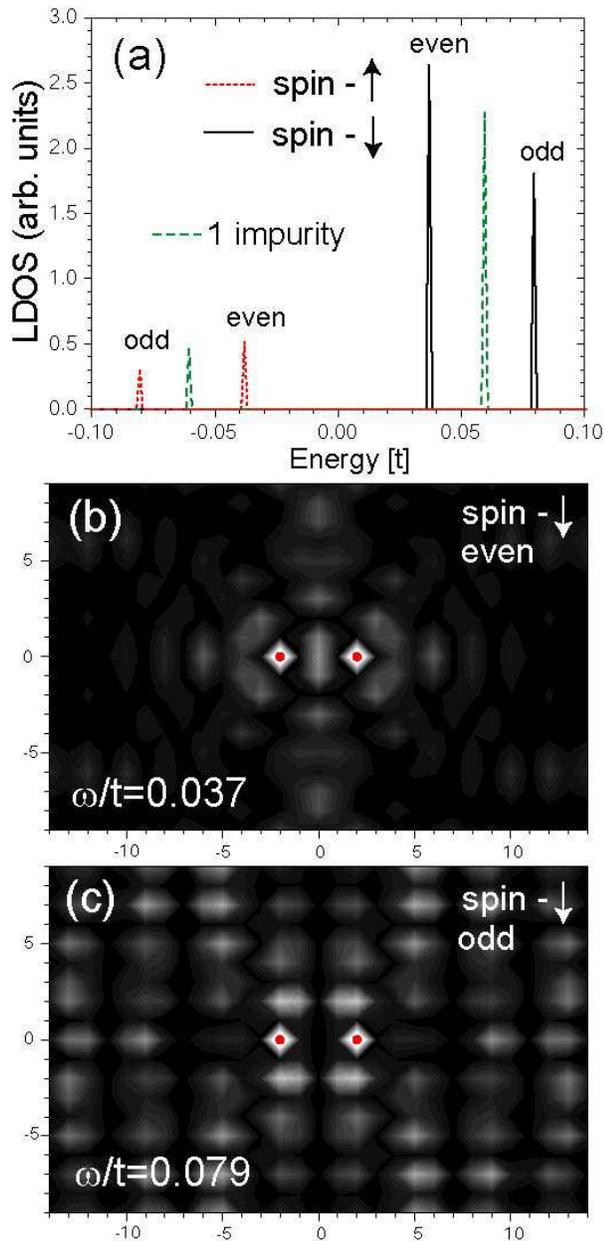,width=8.0cm} \caption{ (Color online). (a)
LDOS at ${\bf r}_{1,2}$ as a function of energy for two impurities
located at ${\bf r}_{1,2}=(\mp 2,0)$ (interimpurity distance $\Delta
r=4$) with $JS/t=2$ (the lattice constant, $a_0$, is set to unity).
Spatial intensity plot of the LDOS at (b) $\omega^e_\downarrow/t =
0.037$, and (c) $\omega^o_\downarrow/t = 0.079$. The filled red
circles denote the positions of the magnetic impurities.}
\label{2imp}
\end{figure}

Before considering the effects of two magnetic impurities on the
local electronic structure of an $s$-wave superconductor, we briefly
review the salient features of the fermionic bound state induced by
a single impurity. For a single magnetic impurity with $\bf{S} \|
\hat{z}$, particle-hole mixing in the superconducting state yields a
wave-function of the induced bound state that possesses a
particle-like and a hole-like component, denoted by $|p,\uparrow
\rangle$ and $|h,\downarrow \rangle$, respectively. Here, $\uparrow,
\downarrow$ denote the opposite spin quantum numbers, $S_z=\pm 1/2$
of the two components. The spectroscopic signature of the bound
state are two peaks in the LDOS, as shown in Fig.~\ref{2imp}(a). The
peak at negative energies arises from the particle-like component of
the wave-function, $|p,\uparrow \rangle$, while the peak at positive
energies represents the hole-like part, $|h,\downarrow \rangle$.
These two peaks are therefore also referred to as the
spin-$\uparrow$ and spin-$\downarrow$ peaks in the LDOS.

We now turn to the case of two magnetic impurities with parallel
spins and $\bf{S} \| \hat{z}$. If the impurities are infinitely far
apart (with interimpurity distance $\Delta r= \infty $), the two
bound states with components $|p,\uparrow, i \rangle$ and
$|h,\downarrow,i \rangle$ ($i=1,2$) are degenerate. This degeneracy
is lifted for $\Delta r < \infty$, since an electron scattered by
one of the impurities is in general also scattered by the second
impurity, resulting in a coupling of the induced bound states. In
order to gain further insight into the nature of this coupling, we
consider a toy model in which the unhybridized bound states of
energy $E_0$ are coupled by a hopping term $D(\Delta r)$, which
depends on the distance, $\Delta r$, between the two impurities.
This coupling, in turn, leads to the formation of even and odd bound
states with energies $E_{e,o}=E_0 \pm D(\Delta r)$ and a
particle-like component of the wavefunction given by $|p,\uparrow
\rangle_{e,o}= (|p,\uparrow, 1 \rangle \pm |p,\uparrow, 2
\rangle)/\sqrt{2}$, and similarly for the hole-like component. As a
result, the bound state peaks in the LDOS are split [by an energy
$\Delta E=E_e-E_o=2D(\Delta r)$], as shown in Fig.~\ref{2imp}(a) for
the case of two impurities, located at ${\bf r}_1=(-2,0)$ and ${\bf
r}_2=(2,0)$ ($\Delta r=4$) with scattering strength $JS/t=2$ (in the
following, we denote the energies of the spin-$\uparrow$ and
spin-$\downarrow$ peaks of the even and odd bound state by
$\omega^{e,o}_{\uparrow,\downarrow}$, respectively). This splitting
of the bound state peaks was also obtained within the
non-self-consistent ${\hat T}$-matrix formalism \cite{Morr03b} (see
also Ref.\cite{Fla00}). In Figs.~\ref{2imp}(b) and (c) we present a
spatial plot of the LDOS at $\omega^e_\downarrow$ and
$\omega^o_\downarrow$, respectively [light (dark) color indicates a
large (small) LDOS, and the filled red circles denote the positions
of the magnetic impurities]. Fig.~\ref{2imp}(c) clearly demonstrates
the odd symmetry of the bound state since the LDOS vanishes at any
point with equal distance to the two impurities.

\begin{figure}[h]
\epsfig{file=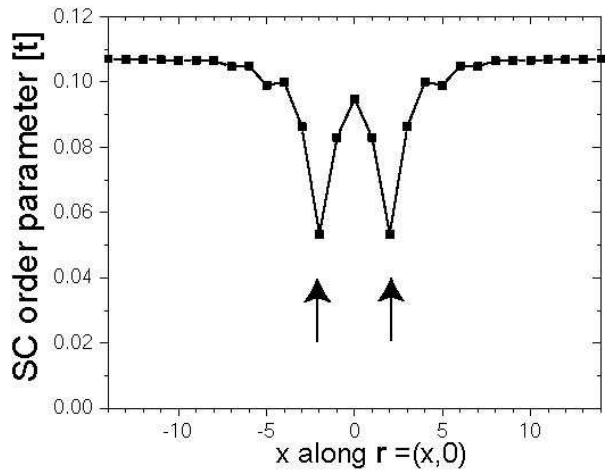,width=8.0cm} \caption{Superconducting order
parameter as a function of position along the line connecting the
two impurities at ${\bf r}_{1,2}=(\mp 2,0)$ considered in
Fig.~\ref{2imp}. The positions of the impurities are denoted by
arrows.} \label{2imp_OP}
\end{figure}

In Fig.~\ref{2imp_OP}, we present the superconducting order
parameter as a function of position for the case of the two
impurities considered in Fig.~\ref{2imp}.
%
%
As expected, we find that the order parameter is suppressed in the
vicinity of the impurities, with the largest suppression at the
impurity sites (indicated by arrows in Fig.~\ref{2imp_OP}). Note,
however, that the order parameter recovers its bulk value within a
few lattice spacing from the impurities. Surprisingly enough, even
for the relatively small interimpurity distance of $\Delta r=4$ the
superconducting order parameter comes close to its bulk value in the
region between the two impurities. We therefore conclude that the
superconducting order parameter relaxes back to its bulk value on a
length scale $\lambda_r$ of a few lattice spacings. This lengthscale
$\lambda_r$ is much shorter than the superconducting coherence
length given by $\xi_c=v_F/\Delta \approx 25 a_0$, where $v_F$ is
the Fermi velocity, and $a_0$ is the lattice spacing. This result is
in agreement with numerical \cite{Sal97} and analytical
\cite{Sch76,Rus69} studies of the superconducting order parameter
near a single impurity, in which the latter identified $\lambda_r$
with the Fermi wavelength $\lambda_F$. In our case, $\lambda_F
\approx 4a_0 \ll \xi_c$.

%
%
\begin{figure}[h]
\epsfig{file=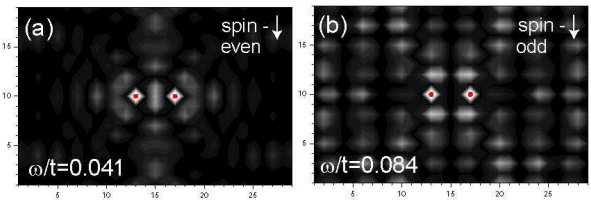,width=8.0cm} \caption{ (Color online). LDOS
for a spatially uniform superconducting order parameter. The
intensity plot of the LDOS is shown at (a) $\omega^e_\downarrow/t =
0.041$, and (b) $\omega^o_\downarrow/t = 0.084$.} \label{CG}
\end{figure}
In order to ascertain the importance of the order parameter
suppression on the frequency and spatial form of the bound states,
we computed the LDOS by constraining the superconducting order
parameter to its spatially homogeneous bulk value. In this case, the
spin-$\uparrow$ and spin-$\downarrow$ peaks of the even and odd
bound state are located at $\omega^e_\uparrow/t=-0.042$,
$\omega^e_\downarrow/t=0.041$, $\omega^o_\uparrow/t=-0.086$ and
$\omega^o_\downarrow/t=0.084$, respectively. In contrast, when a
suppression of the superconducting order parameter is taken into
account within the BdG-formalism, the bound state peaks are located
at $\omega^e_\uparrow/t=-0.038$, $\omega^e_\downarrow/t=0.037$,
$\omega^o_\uparrow/t=-0.081$ and $\omega^o_\downarrow/t=0.079$,
respectively [see Fig.~\ref{2imp}(a)]. Thus the spatial suppression
of the superconducting order parameter leads to a small shift of the
bound state peaks to lower energies (a qualitatively similar effect
was observed near a single magnetic impurity, see
Ref.~\cite{Sal97}). In Figs.~\ref{CG}(a) and (b) we present a
spatial intensity plot of the LDOS at $\omega^e_\downarrow/t=0.041$,
and $\omega^o_\downarrow/t=0.084$, respectively. A comparison of
Figs.~\ref{CG}(a) and (b) with Figs.~\ref{2imp}(b) and (c)
immediately shows that the suppression of the superconducting order
parameter in the vicinity of the impurities does not lead to
perceptible changes in the spatial LDOS pattern of the even and odd
bound states (for better comparison, the intensity scale for all
four figures is the same). Hence, we conclude that characteristic
signatures of quantum interference in the LDOS remain qualitatively
and to a large extent quantitatively unchanged by the suppression of
the order parameter around impurities. This robustness of quantum
interference phenomena arises from the fact that the superconducting
order parameter recovers its bulk value over a length scale which is
typically much shorter than the superconducting coherence length, as
discussed above. This conclusion also supports the validity of the
results pertaining to two impurity interference effects obtained in
Ref.~\cite{Morr03b} within the non-self-consistent ${\hat T}$-matrix
approach.

\begin{figure}[h]
\epsfig{file=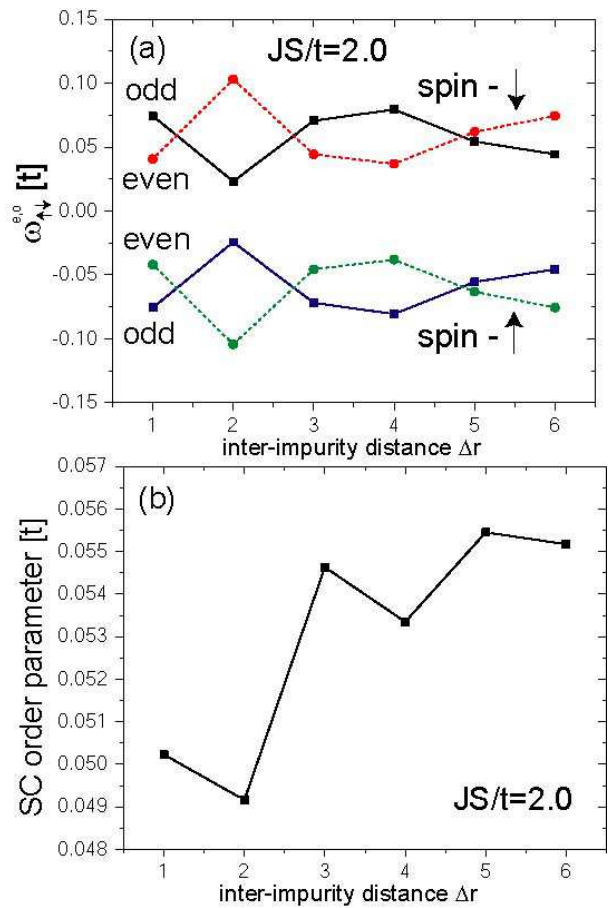,width=8.0cm} \caption{(Color online). (a)
Frequency of the even and odd bound state peaks, and (b) on-site
superconducting order parameter (at ${\bf r}_{1,2}$) as a function
of inter-impurity distance $\Delta r$ for $JS/t=2$.} \label{2imp_dr}
\end{figure}

As the distance between the two impurities is changed, the frequency
of the even and odd bound state peaks, as well as the splitting
between them oscillates \cite{Fla00,Morr03b}, as shown in
Fig.~\ref{2imp_dr}(a).
%
%
The strength of the coupling, $D(\Delta r)$, between the impurity
states can be directly obtained from the frequency splitting between
the even and odd bound states, $\Delta E=E_e-E_o=2D(\Delta r)$. The
results shown in Fig.~\ref{2imp_dr}(a) then imply that the effective
coupling between the impurity states does not only vary in
magnitude, but also changes sign as the distance between the
impurities is changed. One obtains, for example, $D(\Delta r=2)>0$,
while $D(\Delta r=4)<0$. The sign and magnitude of $D(\Delta r)$ is
determined by scattering processes involving both impurities. For
example, consider the wave function of an electron that is scattered
by the impurity at ${\bf r}_1$. The sign and magnitude of its wave
function at ${\bf r}_2$ (and thus those of the effective coupling)
depend (a) on the distance between the impurities due to the $(k_F
r)$-oscillations of the electronic wave-function, and (b) the
scattering phase shift at ${\bf r}_1$, which is determined by the
scattering strength $JS$. As a result of the latter, the relative
splitting between the even and odd bound states in general varies
with $JS$, as demonstrated below. Moreover, if the electron's wave
function vanishes at ${\bf r}_2$, the impurity bound states
associated with each of the two impurities cannot hybridize and
hence remain degenerate \cite{Morr03b}.

Complementary to the oscillations of the bound state frequencies,
the on-site superconducting order parameter at ${\bf r}_{1,2}$ also
oscillates as the distance between the two impurities is increased,
as shown in Fig.~\ref{2imp_dr}(b). Note that a minimum in the
oscillations of the order parameter coincides with a larger
splitting between the even and odd bound states. This is expected
since a larger splitting implies a stronger coupling between the
impurity bound states. At the same time, a larger coupling leads to
a stronger effective scattering strength of the two-impurity system,
and thus to a larger suppression of the superconducting order
parameter.

\begin{figure}[h]
\epsfig{file=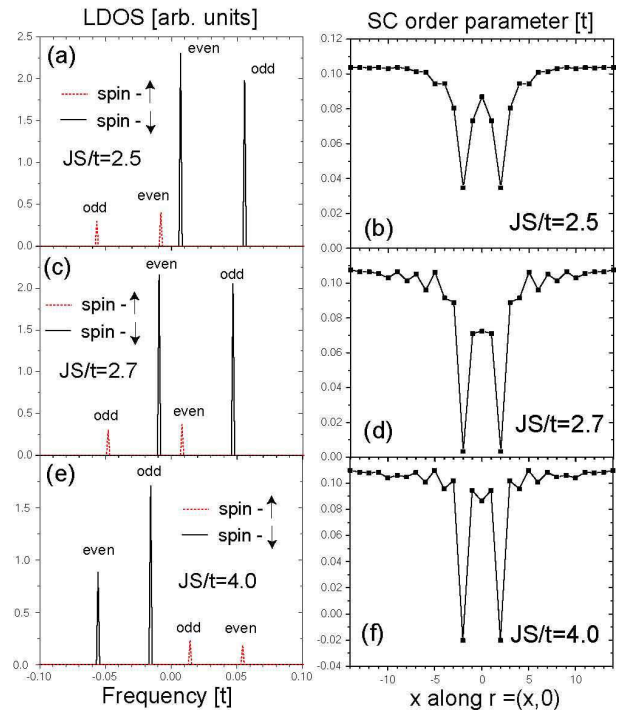,width=8.0cm} \caption{ (Color online). LDOS
at ${\bf r}_{1,2}$ as a function of energy (left column) and
superconducting order parameter along ${\bf r}=(x,0)$ (right column)
for two impurities located at ${\bf r}_1=(-2,0)$ and ${\bf
r}_2=(2,0)$ for (a)(b) $JS/t=2.5$, (c),(d) $JS/t=2.7$, and (e),(f)
$JS/t=4.0$.} \label{LDOS_J}
\end{figure}

With increasing scattering strength, $JS$, of the two magnetic
impurities, the bound states move to lower energies, as follows from
a comparison of the LDOS for $JS/t=2.0$ [see Fig.\ref{2imp}(a)] with
that for $JS/t=2.5$ [see Fig.~\ref{LDOS_J}(a)], and eventually cross
zero energy.
%
%
For the case of a single magnetic impurity in an $s$-wave
superconductor, it was first shown by Sakurai \cite{Sak70} that such
a zero-energy crossing at a critical value, $(JS)_{cr}$, is the
signature of a first order quantum phase transition, in which the
ground state of the superconductor changes from a state with spin
polarization $\langle S_z \rangle=0$ to a state with $\langle S_z
\rangle=\pm 1/2$, depending on the orientation of the magnetic
impurity. Here, the spin polarization is defined as
\begin{equation}
\langle S_z \rangle =\frac{1}{2} \int d^2r \int_{-\infty}^\infty
d\omega \, \left[ N_{\uparrow}({\bf r}, \omega)-N_{\downarrow}({\bf
r}, \omega) \right] n_F(\omega)\ , \label{sz}
\end{equation}
where $N_{\sigma}({\bf r}, \omega)$ is the LDOS of the electrons
with spin $\sigma=\uparrow,\downarrow$, and $n_F$ is the Fermi
distribution function. This phase transition arises from a level
crossing in the superconductor's free energy, ${\cal F}$, resulting
in a discontinuity of $\partial {\cal F}/\partial J$ at the
transition; hence the first order nature of the transition (for a
more detailed discussion see Ref.~\cite{Balatsky_RMP}). For a single
magnetic impurity, the transition also possesses a characteristic
signature in the on-site order parameter which exhibits a
discontinuous decreases and changes sign at $(JS)_{cr}$, thus
undergoing a $\pi$-phase shift in comparison to its bulk value.

For the case of two magnetic impurities, quantum interference leads
to the emergence of two quantum phase transitions at $(JS)_{cr,1}$
and $(JS)_{cr,2}$, corresponding to the transitions $\langle S_z
\rangle=0 \rightarrow -1/2$ and $\langle S_z \rangle=-1/2
\rightarrow -1$, respectively, of the ground state's spin
polarization. Specifically, the even bound state crosses zero energy
at $(JS)_{cr,1}/t \approx 2.54$, such that for $JS>(JS)_{cr,1}$, the
spin-$\uparrow$ (spin-$\downarrow$) peak of the even bound state is
hole-like (particle-like), as shown in Fig.~\ref{LDOS_J}(c) for
$JS/t=2.7$. However, while in the non-self-consistent ${\hat
T}$-matrix approach, the frequency of the even bound state,
$\omega^e_{\uparrow,\downarrow}$ evolves continuously as a function
of $JS$ and reaches zero at the transition \cite{Morr03b}, within
the BdG formalism, $\omega^e_{\uparrow,\downarrow}$ crosses
zero-energy discontinuously at $(JS)_{cr}$, similar to the
transition associated with a single magnetic impurity \cite{Sal97}.
Specifically, for $(JS)_{cr,1}-0^+$ one finds
$\omega^e_{\uparrow,\downarrow}/t= \mp 0.005$, while for
$(JS)_{cr,1}+0^+$ one has $\omega^e_{\uparrow,\downarrow}/t=\pm
0.001$ [the upper (lower) sign corresponds to the spin-$\uparrow$
(spin-$\downarrow$) peak]. Interestingly enough, we find that the
energy of the odd bound state does not exhibit a discontinuity at
$(JS)_{cr,1}$. A similar discontinuous zero-energy crossing of the
bound state energies is found at all phase transitions discussed
below. The second quantum phase transition (with $\langle S_z
\rangle=-1/2 \rightarrow -1$) occurs at $(JS)_{cr,2}/t=3.69$ when
the odd bound state crosses zero energy, such that for
$JS>(JS)_{cr,2}$, all spin-$\uparrow$ (spin-$\downarrow$) peaks are
now located at positive (negative) energies [see
Fig.~\ref{LDOS_J}(e) for $JS=4.0t>(JS)_{cr,2}$].

\begin{figure}[!h]
\epsfig{file=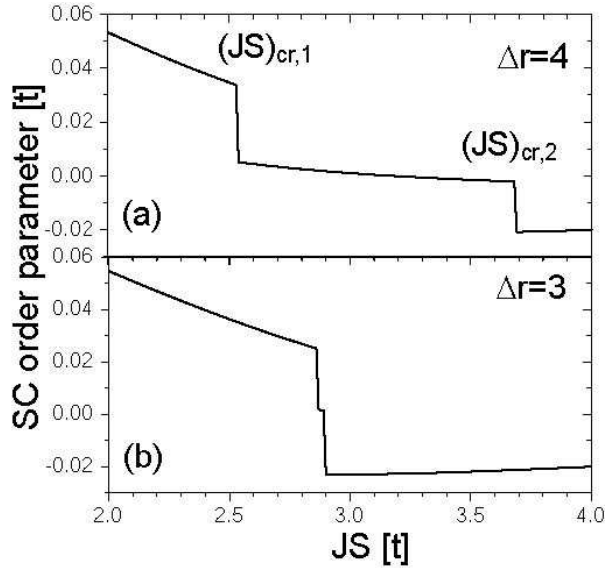,width=8.0cm} \caption{On-site superconducting
order parameter as a function of $JS$ for two impurities located at
(a) ${\bf r}_{1,2}=(\mp 2,0)$ with $\Delta r=4$ (the critical values
$(JS)_{cr,j}$ are indicated by arrows) and (b) ${\bf r}_1=(-1,0)$
and ${\bf r}_2=(2,0)$ with $\Delta r=3$.} \label{SCgap_J}
\end{figure}

In Figs.~\ref{LDOS_J}(b), (d) and (f) we plot the superconducting
order parameter as a function of position for the three values of
$JS$ considered above. We find that the two phase transitions are
accompanied by discontinuous changes of the on-site superconducting
order parameter at ${\bf r}_{1,2}$, as follows directly from
Fig.~\ref{SCgap_J}(a) where we plot the on-site order parameter as a
function of $JS$.
%
%
Note, however, that while the on-site order parameter is reduced at
$(JS)_{cr,1}$, it remains positive and does not exhibit a
$\pi$-phase shift, in qualitative contrast to the case of a single
magnetic impurity \cite{Balatsky_RMP,Sak70,Sal97}. For
$JS>(JS)_{cr,1}$ the superconducting order parameter decreases
continuously with increasing $JS$, and crosses zero at $JS/t \approx
3.15$. This zero-crossing of the on-site order parameter does not
lead to any signature in the LDOS. At $(JS)_{cr,2}$, the
superconducting order parameter at ${\bf r}_{1,2}$ exhibits a second
discontinuity. Note, however, that in contrast to the discontinuous
change at $(JS)_{cr,1}$, where the magnitude of the superconducting
order parameter decreases, its magnitude increases at $(JS)_{cr,2}$.
Interestingly enough, the suppression of the superconducting order
parameter becomes spatially more confined when the system undergoes
a phase transition. For example, the superconducting order parameter
at ${\bf r}_{nn}=(3,0)$, i.e., one site away from one of the
impurities, is much less reduced from its bulk value for
$JS>(JS)_{cr,1}$ [Fig.~\ref{LDOS_J}(d)] than for $JS<(JS)_{cr,1}$
[Fig.~\ref{LDOS_J}(b)]. Moreover, for $JS>(JS)_{cr,2}$ the order
parameter at ${\bf r}_{nn}$ as well as between the two impurities at
${\bf r}=(0,0)$ is quite close to its bulk value [see
Fig.~\ref{LDOS_J}(f)], in contrast to the results shown in
Fig.~\ref{LDOS_J}(b) for $JS<(JS)_{cr,1}$.

\begin{figure}[h]
\epsfig{file=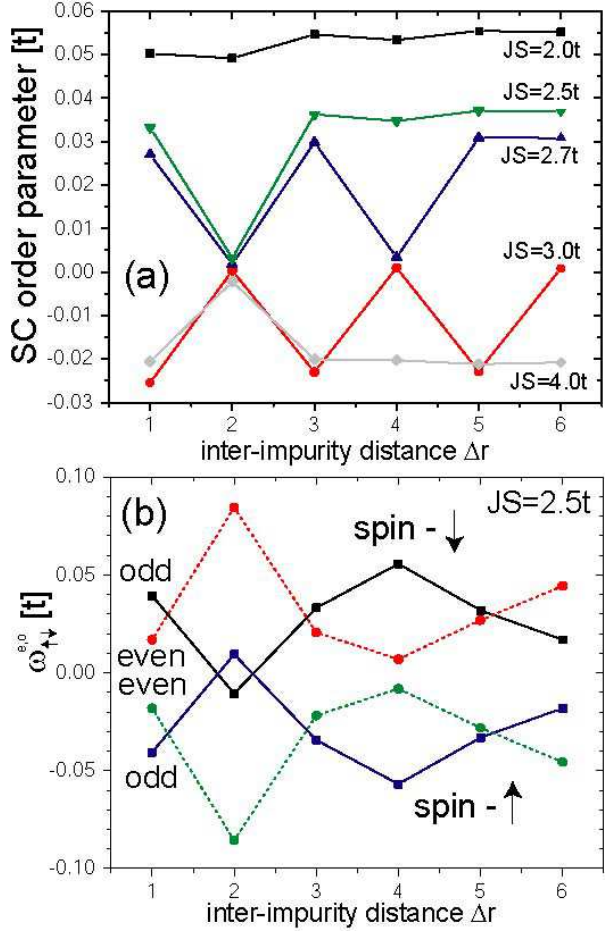,width=8.0cm} \caption{(Color online). (a)
Superconducting order parameter, $\Delta_{{\bf r}_{1,2}}$ at ${\bf
r}_{1,2}$ as a function of interimpurity distance, $\Delta r$, for
several values of $JS/t$. (b) Frequency of the even and odd bound
state peaks as a function of inter-impurity distance $\Delta r$ for
$JS/t=2.5$.} \label{SCgap_dr}
\end{figure}

For all interimpurity distances we considered, the dependence of the
LDOS and the superconducting order parameter on the scattering
strength is similar to the one discussed above. However, the
splitting between $(JS)_{cr,1}$ and $(JS)_{cr,2}$ depends strongly
on the interimpurity distance. For example, for $\Delta r=3$, the
two phase transitions occur at $(JS)_{cr,1}/t=2.87$ and
$(JS)_{cr,2}/t=2.9$, as shown in Fig.~\ref{SCgap_J}(b). The
splitting between the two critical values is thus considerably
smaller than for the case $\Delta r=4$ shown in
Fig.~\ref{SCgap_J}(a). We note, however, that while the splitting
between the two critical values is small, they are nevertheless
reduced from the critical value of a single magnetic impurity given
by $(JS)_{cr}/t=3.09$. At the same time, we find that the energy
splitting between the odd and even bound states is considerably
smaller for $\Delta r=3$ than for $\Delta r=4$. In particular, for
$\Delta r=3$ the frequency splitting between the even and odd bound
state peaks at $(JS)_{cr,1}$ is $\Delta \omega \approx 0.0056t$,
while for $\Delta r=4$ at $(JS)_{cr,1}$, the frequency splitting is
about one order of magnitude larger with $\Delta \omega \approx
0.056t$. We therefore conclude that the splitting of the critical
values of $JS$, similar to the splitting of the even and odd bound
state energies depends on the effective coupling between the two
impurities. Hence, $(JS)_{cr,1,2}$ are functions of $\Delta r$. In
this regard, it is interesting to note that with increasing $JS$,
the frequency splitting between the odd and even bound states
increases for $\Delta r=4$, but decreases for $\Delta r=3$, as can
be seen from a comparison of Fig.~\ref{2imp_dr}(a) and
Fig.~\ref{SCgap_dr}(b). This effect directly reflects the dependence
of the scattering phase shift on $JS$, as discussed above.

In Fig.~\ref{SCgap_dr}(a) we present the superconducting order
parameter at ${\bf r}_{1,2}$ for several values of $JS$ as a
function of inter-impurity distance, $\Delta r$.
%
%
For $JS/t=2$, the order parameter exhibits only weak Friedel-like
oscillations when the distance between the impurities is varied (see
also Fig.~\ref{2imp_OP}). In contrast, for $JS/t=2.7$, the on-site
superconducting order parameter, $\Delta_{{\bf r}_{1,2}}$,
oscillates much more strongly when the interimpurity distance is
changed, for example, from $\Delta_{{\bf r}_{1,2}}=0.030t$ at
$\Delta r=3$ to $\Delta_{{\bf r}_{1,2}}=0.003t$ at $\Delta r=4$. The
reason for this strong variation becomes clear when one considers
Fig.~\ref{SCgap_J}. For $\Delta r=3$, one finds
$(JS)_{cr,1}/t>JS/t=2.7$, implying that the spin polarization of the
superconductor is $\langle S_z \rangle=0$. In contrast, for $\Delta
r=4$ one has $(JS)_{cr,1}/t<JS/t=2.7<(JS)_{cr,2}/t$, yielding
$\langle S_z \rangle=-1/2$.  In other words, for $JS/t=2.7$ the
superconductor undergoes a first order phase transition when the
interimpurity distance is changed from $\Delta r=3$ with spin
polarization $\langle S_z \rangle=0$ to $\Delta r=4$ where $\langle
S_z \rangle=-1/2$. As a result, the superconducting order parameter
varies strongly between $\Delta r=3$ and $\Delta r=4$, as shown in
Fig.~\ref{SCgap_dr}(a). This effect directly reflects the dependence
of $(JS)_{cr,1(2)}$ on $\Delta r$, as discussed above.

While for $JS/t=2.7$, the spin polarization of the superconductor's
ground state changes between $\langle S_z \rangle=0$ and $\langle
S_z \rangle=-1/2$ , we find that for $JS/t=3.0$, its spin
polarization varies between $\langle S_z \rangle=-1/2$ [for $\Delta
r=2,4,6$ where $(JS)_{cr,1}/t<JS/t=3.0<(JS)_{cr,2}/t$] and $\langle
S_z \rangle=-1$ [for $\Delta r=1,3,5$ where
$(JS)_{cr,2}/t<JS/t=3.0$]. Finally, for $JS/t=4.0$, the behavior of
the superconducting order parameter is complementary to that at
$JS/t=2.5$, in that the superconductor's spin polarization is
$\langle S_z \rangle=-1/2$ for $\Delta r=2$, and $\langle S_z
\rangle=-1$ for $\Delta r \not =2$ (while $\langle S_z \rangle=0$
for $JS/t=2.5$ and $\Delta r \not =2$).

In order to explore the interplay between the position of the bound
state peaks in the LDOS and the superconducting order parameter
further, we plot in Fig.~\ref{SCgap_dr}(b) the energies of the even
and odd bound states as a function of inter-impurity distance for
$JS/t=2.5$. For $\Delta r=2$, the spin polarization of the
superconductor is given by $\langle S_z \rangle=-1/2$, since the odd
bound state has crossed zero energy. At the same time, the
superconducting order parameter at ${\bf r}_{1,2}$ is significantly
reduced [see Fig.~\ref{SCgap_dr}(a)]. In contrast, for all other
interimpurity distances, no zero-energy crossing of the bound state
peaks is observed, and the superconducting order parameter at ${\bf
r}_{1,2}$ is significantly larger than that for $\Delta r=2$. Both
results imply that for $\Delta r \not =2$, one has $\langle S_z
\rangle=0$.

\section{Three magnetic impurities}
\label{3impurities}

We next consider the case of three magnetic impurities, for which
the superconducting order parameter exhibits an interesting behavior
at the first order phase transitions. Specifically, we study three
impurities with parallel spins ($\bf{S} \| \hat{z}$) that are
located at ${\bf r}_{1,3}=(\mp x,0)$ and ${\bf r}_{2}=(0,0)$.
Quantum interference of scattered electrons again lifts the
degeneracy of the fermionic bound states associated with each of the
impurities, leading to the emergence of six peaks in the LDOS. In
order to gain insight into the symmetry of the hybridized bound
states, we assume that the {\it unhybridized} bound states of energy
$E_0$ are coupled to the nearest neighbor (next nearest neighbor)
state via a hopping term $K$ ($K^\prime$). The resulting energies of
the three hybridized bound states are then given by
\begin{eqnarray*}
E_{odd}&=&E_0-K^\prime \\
E_{\pm}&=&E_0 +\frac{K^\prime}{2} \pm
\sqrt{\frac{(K^\prime)^2}{4}+2K^2}
\end{eqnarray*}
The wavefunction with energy $E_{odd}$ is given by $\Psi_{odd}=
(\psi_1-\psi_3)/\sqrt{2}$ where $\psi_i$ is the wavefunction of the
unhybridized fermionic bound state induced by the impurity at ${\bf
r}_i$. Note that the wavefunction $\Psi_{odd}$ is odd and thus
vanishes at any point in space with equal distance to ${\bf r}_1$
and ${\bf r}_3$. Thus  the odd bound state vanishes in particular at
${\bf r}_2$, which, as discussed below, is important for the
behavior of the superconducting order parameter at the first order
phase transitions. The wavefunctions with energies $E_{\pm}$ for the
case $K^\prime=0$ are given by $\Psi_{\pm}=\sqrt{2}(\psi_1 \pm
\psi_2/\sqrt{2}-\psi_3)/\sqrt{5}$, respectively (we omit their forms
for $ K^\prime \not = 0$ since these are quite cumbersome and
irrelevant for the discussion below). Note that the energy of the
odd bound state always lies between those of the
$\Psi_{\pm}$-states, with the energy difference between the
$\Psi_{\pm}$-states and the $\Psi_{odd}$-state given by
\begin{equation}
\Delta E_\pm=E_\pm -E_{odd}=\frac{3 K^\prime}{2} \pm
\sqrt{\frac{(K^\prime)^2}{4}+2K^2} \ . \label{Ediff}
\end{equation}

\begin{figure}[t]
\epsfig{file=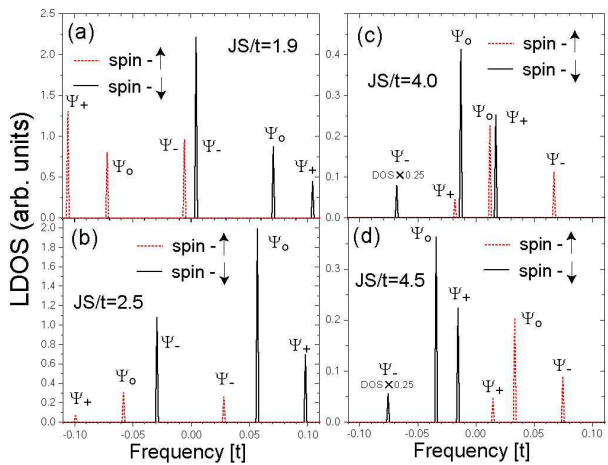,width=8.0cm} \caption{ (Color online). LDOS
as a function of frequency at ${\bf r}_{1,3}$ for the case of three
impurities with interimpurity distance $\Delta r=2$  and four
different values of $JS$: (a) $JS/t=1.9$, (b) $JS/t=2.5$,(c)
$JS/t=4.0$, and (d) $JS/t=4.5$. The results in (a) were obtained for
a lattice with $N_x=34$ and $N_y=22$, those in (b)-(d) for a lattice
with $N_x=28$ and $N_x=18$ (see Ref.~\cite{com1}).}
\label{LDOS_3imp_r2}
\end{figure}

Next, we consider the concrete case of three impurities located at
${\bf r}_{1,3}=(\mp 2,0)$ and ${\bf r}_2=(0,0)$ with nearest
neighbor distance $\Delta r=2$. The resulting LDOS as a function of
frequency at ${\bf r}_{1,3}$ for four different values of $JS$ is
shown in Fig.~\ref{LDOS_3imp_r2}.
%
%
As discussed above, one finds six peaks in the LDOS, three peaks
each arising from the components of the hybridized bound state wave
functions with $S_z=+1/2$ (spin-$\uparrow$ peaks) and $S_z=-1/2$
(spin-$\downarrow$ peaks). For small $JS$
[Fig.~\ref{LDOS_3imp_r2}(a)], all spin-$\downarrow$
(spin-$\uparrow$) peaks are hole-like (particle-like) and thus
located at $\omega>0$ ($\omega<0$). With increasing $JS$, the peaks
move towards zero energy, and a phase transition occurs when the
first bound state crosses zero energy at $(JS)_{cr,1}/t=1.98$. At
this first order transition, the spin polarization of the
superconductors changes from $\langle S_z \rangle =0$ to $\langle
S_z \rangle =-1/2$. For $JS=2.5t>(JS)_{cr,1}$ [see
Fig.~\ref{LDOS_3imp_r2}(b)] one of the spin-$\downarrow$
(spin-$\uparrow$) peaks is located at negative (positive) energies.

\begin{figure}[h]
\epsfig{file=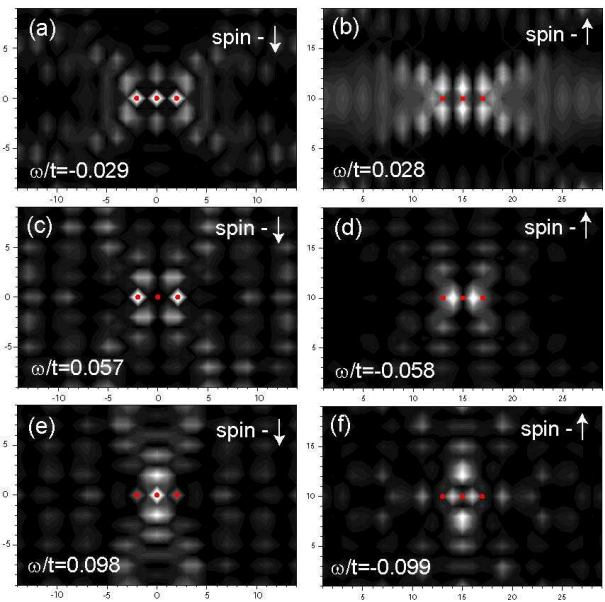,width=8.0cm} \caption{ (Color online).
Spatial intensity plot of the LDOS for three impurities with
$JS/t=2.5$ at the frequency of the spin-$\downarrow$ peaks and
spin-$\uparrow$ peaks of the $\Psi_-$-state [(a) and (b)], the
$\Psi_{odd}$-state [(c) and (d)],
 and the $\Psi_+$-state [(e) and (f)], respectively.} \label{LDOSint_3imp_r2}
\end{figure}

The spatial pattern of the three bound states at the frequencies of
the spin-$\uparrow$ and spin-$\downarrow$ peaks is shown in
Fig.~\ref{LDOSint_3imp_r2}.
%
%
The bound state whose spatial pattern is shown in
Fig.~\ref{LDOSint_3imp_r2}(c) and (d) is easily identified as the
odd bound state with wavefunction $\Psi_{odd}$. In order to identify
the $\Psi_\pm$-bound states, we use Eq.(\ref{Ediff}) and the results
presented in Figs.~\ref{2imp_dr}(a) and \ref{SCgap_dr}(b). Since
$K=D(\Delta r=2)>0$, and $K^\prime=D(\Delta r=4)<0$, we obtain from
Eq.(\ref{Ediff}) that $|\Delta E_-|>|\Delta E_+|$. An inspection of
the LDOS in Fig.~\ref{LDOS_3imp_r2} then allows us to identify the
bound state whose spatial pattern is shown in
Figs.~\ref{LDOSint_3imp_r2}(a) and (b) as the $\Psi_-$-state [which
crosses zero energy at $(JS)_{cr,1}$], while the bound state shown
in Figs.~\ref{LDOSint_3imp_r2}(e) and (f) is the $\Psi_+$-state.
This result is also supported by our analysis of the LDOS$\sim
|\Psi_{\pm}|^2$ based on the value of $\Psi_{\pm}$ at ${\bf
r}_{1,2}$. Finally, we note that the spatial structure of these
bound states remains almost unchanged as $JS$ is increased and the
superconductor undergoes a first order transition.

\begin{figure}[h]
\epsfig{file=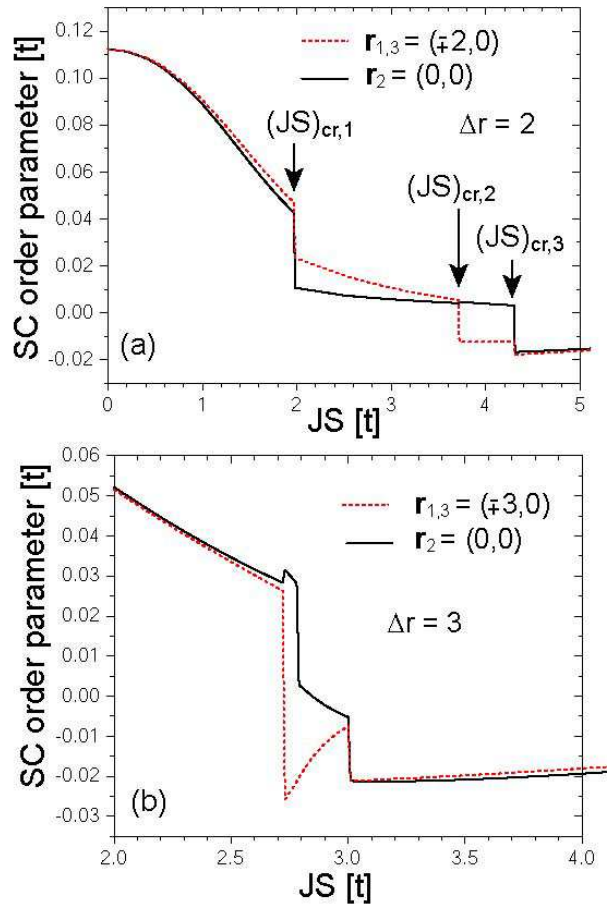,width=8.0cm} \caption{ (Color online).
On-site superconducting order parameter at ${\bf r}_{1,2,3}$ as a
function of $JS$ for three impurities with nearest neighbor distance
(a) $\Delta r=2$ (the critical values $(JS)_{cr,j}$ are indicated by
arrows) and (b) $\Delta r=3$.} \label{SCOP_3imp_J}
\end{figure}

The zero energy crossing of the $\Psi_-$-state at $(JS)_{cr,1}$ is
accompanied by a discontinuous jump in the superconducting order
parameter at all three impurity sites, as shown in
Fig.~\ref{SCOP_3imp_J}(a).
%
%
While the superconducting order parameter is reduced at the
transition, it remains positive, and does not undergo a $\pi$-phase
shift. When $JS$ is further increased, the odd bound state crosses
zero energy at $(JS)_{cr,2}/t=3.72$, and the superconductor
undergoes a first order phase transition to a state with $\langle
S_z \rangle =-1$ [a spatial intensity plot of the LDOS at frequency
of the odd bound state is shown in Figs.~\ref{LDOSint_3imp_r2}(c)
and (d)]. For $JS=4.0t>(JS)_{cr,2}$, two spin-$\downarrow$
(spin-$\uparrow$) peak are located at $\omega<0$ ($\omega>0$) [see
Fig.~\ref{LDOS_3imp_r2}(c)]. Since the odd bound state vanishes at
${\bf r}_2$, we find that the superconducting order parameter at
${\bf r}_2$ remains unaffected by the zero energy crossing, while
that at ${\bf r}_{1,3}$ changes discontinuously and undergoes a
$\pi$-phase shift, as shown in Fig.~\ref{SCOP_3imp_J}(a). Finally,
at $(JS)_{cr,3}/t=4.31$ the $\Psi_+$-bound state crosses zero
energy, and the superconductor undergoes a first order phase
transition to a state with a spin polarization $\langle S_z \rangle
= -3/2$ [the spatial LDOS intensity of the $\Psi_+$ bound state is
shown in Figs.~\ref{LDOSint_3imp_r2}(e) and (f)]. All
spin-$\downarrow$ (spin-$\uparrow$) peaks are now located at
$\omega<0$ ($\omega>0$), as shown in Fig.~\ref{LDOS_3imp_r2}(c) for
$JS=4.5t>(JS)_{cr,3}$. At $(JS)_{cr,3}$, the superconducting order
parameter at all three impurity sites again changes discontinuously,
and increases in magnitude, as shown in Fig.~\ref{SCOP_3imp_J}(a).
However, only the order parameter at ${\bf r}_2$ changes sign at
this transition.

\begin{figure}[h]
\epsfig{file=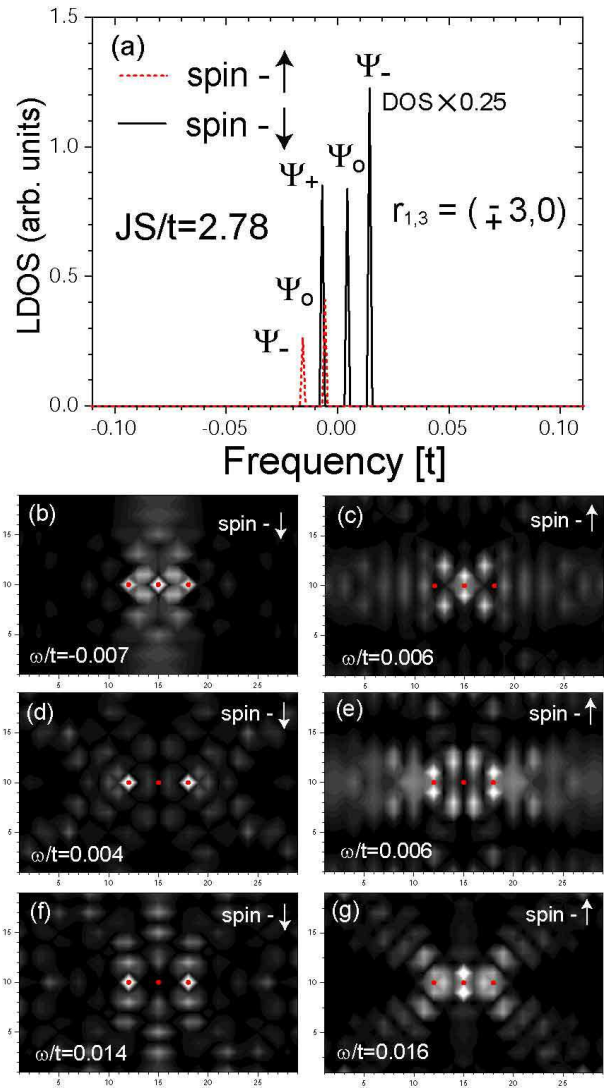,width=8.0cm} \caption{(Color online). LDOS
for three impurities with $\Delta r=3$ and $JS/t=2.78>(JS)_{cr,1}$.
(a) LDOS at ${\bf r}_{1,3}$ as a function of frequency. Spatial
intensity plot of the LDOS at the frequency of the spin-$\downarrow$
peaks and spin-$\uparrow$ peaks of the $\Psi_+$-state [(b) and (c)],
the $\Psi_{odd}$-state [(d) and (e)], and the $\Psi_-$-state [(f)
and (g)], respectively.} \label{LDOS_3imp_r3}
\end{figure}

The behavior of the superconducting order parameter at the first
order phase transitions is different from the above scenario when
the nearest neighbor interimpurity distance is increased to $\Delta
r=3$ and the impurities are located at ${\bf r}_{1,3}=(\mp 3,0)$ and
${\bf r}_2=(0,0)$. In this case, the three phase transitions occur
in a much smaller range of $JS$ than for $\Delta r=2$, with
$(JS)_{cr,1}/t=2.73$, $(JS)_{cr,2}/t=2.79$ and $(JS)_{cr,3}/t=3.01$.
As discussed above, this indicates a weaker coupling between the
unhybridized bound states of the impurities, and we therefore also
expect a smaller frequency splitting between the peaks of the three
bound states in the LDOS. This smaller frequency splitting is
directly evident from a comparison of the LDOS for $\Delta r=3$ and
$JS/t=2.78>(JS)_{cr,1}$ shown in Fig.~\ref{LDOS_3imp_r3}, with the
LDOS for $\Delta r=2$ shown in Fig.~\ref{LDOS_3imp_r2}.
%
%
Moreover, the fact that the energy splittings between the peaks,
$|\Delta E_{\pm}|$, is almost identical, implies that the effect of
$K^\prime = D(\Delta r=6)$ can be neglected in comparison to that of
$K = D(\Delta r=3)$. A more detailed analysis of $\Psi_\pm$ then
shows that for $\Delta r=3$, the relative energies of the $\Psi_+$
and $\Psi_-$ bound states are reversed in comparison to the case
$\Delta r=2$: the peaks of the $\Psi_-$ bound state are now located
at higher energies than those of the $\Psi_+$ bound state. Thus the
$\Psi_+$ bound state crosses zero energy at $(JS)_{cr,1}$, the
$\Psi_{odd}$ bound state at $(JS)_{cr,2}$, and the $\Psi_-$ bound
state at $(JS)_{cr,3}$. Interestingly enough, the LDOS shown in
Fig.~\ref{LDOS_3imp_r3}(a) at ${\bf r}_{1,3}$ only possesses five
peaks. The reason for the ``missing" peak becomes evident when one
considers the spatial form of the LDOS at the bound state
frequencies. Specifically, the LDOS of the $\Psi_+$  bound state at
the frequency of the spin-$\uparrow$ peak vanishes at ${\bf
r}_{1,3}$, as shown in Fig.~\ref{LDOS_3imp_r3}(c), which explains
the absence of the sixth bound state peak in
Fig.~\ref{LDOS_3imp_r3}(a). Note that the LDOS at the
spin-$\downarrow$ peak of the $\Psi_+$ bound state does not vanish
at ${\bf r}_{1,3}$ [see Fig.~\ref{LDOS_3imp_r3}(b)]. This spatial
form of the $\Psi_+$  bound state possesses an interesting
consequence for the behavior of the superconducting order parameter
at $(JS)_{cr,1}$, as shown in Fig.~\ref{SCOP_3imp_J}(b).
Specifically, at $(JS)_{cr,1}$ the superconducting order parameter
at ${\bf r}_2$ is reduced, while the order parameter at ${\bf
r}_{1,3}$ is slightly enhanced. This behavior of the superconducting
order parameter is quite different from that at $(JS)_{cr,3}$, when
the $\Psi_-$  bound state crosses zero energy. Since neither the
LDOS of its spin-$\downarrow$ peak nor of its spin-$\uparrow$ peak
vanishes at ${\bf r}_{1,3}$ (or ${\bf r}_{2}$), we find that at
$(JS)_{cr,3}$, the order parameter at all three impurity sites
decreases discontinuously. Moreover, when the LDOS of both the
spin-$\downarrow$ and spin-$\uparrow$ peak vanishes at an impurity
position, as is the case with  $\Psi_{odd}$ bound state at ${\bf
r}_2$ [see Figs.~\ref{LDOS_3imp_r3}(d) and (e)], the superconducting
order parameter at this impurity site is unaffected by the zero
energy crossing of the bound state, as follows from
Figs.~\ref{SCOP_3imp_J}(a) and (b). Thus the vanishing of the LDOS
of only one of the spin-$\uparrow$ or spin-$\downarrow$ peaks at an
impurity site can lead to a discontinuous increase in the
superconducting order parameter at the phase transition, as
discussed above.

\section{Conclusions}
\label{conclusion}

In conclusion, we have studied quantum interference phenomena in
impurity structures consisting of two or three magnetic impurities
located on the surface of an $s$-wave superconductor. By using a
self-consistent BdG formalism \cite{Gen89}, we simultaneously
investigated how quantum interference affects (a) the formation of
impurity bound states in the LDOS, and (b) the spatial form of the
superconducting order parameter.

We obtained three important results. First, we find that the
characteristic signatures of quantum interference in the LDOS and in
the superconducting order parameter are coupled. In particular, the
larger the splitting between the hybridized bound states in the
LDOS, the stronger is the suppression of the superconducting order
parameter at the impurity sites. As a result, both the frequency of
the bound states as well as the on-site order parameter exhibit
Friedel-like oscillations as the distance between impurities is
changed. Second, the suppression of the superconducting order
parameter around impurities does not qualitatively change the
signatures of quantum interference in the LDOS, thus demonstrating
the robustness of quantum interference phenomena. The physical
origin of this robustness lies in the fact that, while the
superconducting order parameter is suppressed by impurities, it
recovers its bulk value over a length scale which is set by the
Fermi wavelength, $\lambda_F$, and not by the superconducting
coherence length, which is typically much larger than $\lambda_F$.
Third, we show that quantum interference leads to multiple first
order quantum phase transitions in the superconductor, which are
accompanied by a change in the spin polarization of the
superconductor's ground state. The superconductor can be tuned
through these transitions by increasing the impurities' scattering
strength or by changing the interimpurity distance, the latter being
the experimentally more relevant tuning. These quantum transitions
exhibit several characteristic features that are qualitatively
different from the ones of a phase transition associated with a
single magnetic impurity. In particular, the superconducting order
parameter does not necessarily decrease discontinuously or undergo a
$\pi$-phase shift, and, in certain cases, can even be enhanced at
the transition. The difference between the critical values of $JS$
for which these multiple transitions occur is determined by the
hybridization strength, and is thus directly linked to the splitting
between the impurity bound state energies. The tuning of the
superconductor's spin polarization by changing the interimpurity
distance potentially possesses important applications in the field
of spin electronics \cite{Wolf01} and quantum information technology
\cite{Kane98} as it might lead to the creation of new types of
quantum qubits.

Finally, we note that the relative orientation of the impurities'
magnetic moments is determined in general by the residual
interactions between the impurities, whose consideration is beyond
the scope of this study. In case where the dominant contribution to
the inter-impurity interaction arises from an RKKY-type process, the
interesting question arises to what extent the interaction is
affected by the formation of the hybridized impurity states. This
question will be addressed in some future work.

\section{Acknowledgement}

We would like to thank J.C. Davis, A. Kapitulnik and A. Yazdani for
stimulating discussions. D.K.M. acknowledges financial support by the
Alexander von Humboldt
Foundation, the National Science Foundation under Grant No.
DMR-0513415 and the U.S. Department of Energy under Award No.
DE-FG02-05ER46225.


\begin{thebibliography}{99}
\bibitem{com2} However, for two magnetic impurities with
$\alpha=\pi$, no hybridization of the impurity bound states takes
place \cite{Morr03b}.
\bibitem{com1} We find that for the 3-impurity system considered in
Fig.~\ref{LDOS_3imp_r2} and a lattice with $N_x=28$ and $N_x=18$,
the bound state peaks highest in energy are too close to the
superconducting coherence peaks to be properly resolved for
$JS<(JS)_{cr,1}$. Therefore, we considered a slightly larger lattice
with $N_x=34$ and $N_y=22$ for $JS<(JS)_{cr,1}$ [see
Fig.~\ref{LDOS_3imp_r2}(a)] for which all three sets of bound state
peaks are clearly resolved.
\end{thebibliography}
\end{document}